\documentclass[12pt]{article}

\parskip=\medskipamount            % space between paragraphs (LaTeX)
\def\7#1#2{\mathop{\null#2}\limits^{#1}}        % puts #1 atop #2
\def\ast{\displaystyle *}

\def\beee{\begin{equation}}
\def\eeee{\end{equation}}

\begin{document}
\bibliographystyle{unsrt}
 
\begin{center}
\textbf{A SCHEMATIC MODEL OF\\ GENERATIONS OF QUARKS AND LEPTONS}\\
[5mm]
O.W. Greenberg\footnote{email address, owgreen@physics.umd.edu.}\\
{\it Center for Theoretical Physics\\
Department of Physics\\
University of Maryland\\
College Park, MD~~20742-4111, USA,\\
Helsinki Institute of Physics\\
University of Helsinki\\
FIN-00014 Helsinki, Finland,\\
and Dublin Institute for Advanced Studies\\
Dublin 4, Ireland\\
University of Maryland Preprint PP-06-010}\\
~\\
\end{center}

\begin{abstract}

We propose a model of generations that has exactly three generations. This model has
several attractive features: There is a simple mechanism to produce
the CKM quark mixings and their neutrino analogs. There are definite predictions 
for particles of much higher mass
than the quarks and leptons of the standard model, including the prediction of
two-body decays without missing mass for the higher mass particles. There is 
a natural dark matter particle. A discussion of masses of the quarks
suggests that the large mass
differences between the generations could have a qualitative explanation, and there could be 
a simple way to
make the $u$-$d$ mass difference negative, but the $c$-$s$ and $t$-$b$ mass
differences positive. The model also suggests a completely new scenario for the production,
mass and decays of the Higgs meson that will be analysed in a separate paper.

\end{abstract}

\section{Introduction}
The purpose of this paper is to introduce a simple composite model for generations.
In this model (a) exactly three generations occur naturally,
(b) the CKM mass and mixing matrix for quarks and the
analog for neutrinos have a simple mechanism, (c) there is a definite prediction for 
higher-mass particles with unique two-body decays with no missing energy, (d) the 
arbitrary tunings of
Yukawa couplings that are required by the standard model to fit the quark and lepton
masses are replaced by a dynamical
mechanism, (e) the Higgs phenomenon enters in a new way with new consequences for
the production and decays of the Higgs meson,
(f) there is a natural dark matter candidate,
(g) the large mass difference between
successive generations can have a qualitative explanation, and the up-down mass difference
reverses sign between the first and the last two generations in a natural way 
(according to an exploratory calculation of the masses of
quarks and leptons in the Appendix).

The pattern of masses and mixings of quarks and leptons is a central problem of
elementary particle physics. Analogous problems, such as the spectrum of atoms,
the spectrum of nuclei and the spectrum of hadrons, have been solved by composite models;
for atoms, the Bohr atom and its extensions; for nuclei, the proton-neutron model
and its elucidations; for hadrons, the quark model~\cite{gel,zwe}. Early suggestions that quarks
and leptons could be composite were given in~\cite{ram,owgcan,owgq,har,owgsuch,cas},
but no model has been successful up to now. Most theorists who work in
elementary particle physics have abandoned the composite model approach.
Some introduce new symmetries that
relate the generations (families), for example~\cite{app}.
Most of the activity in elementary particle theory at present searches for new
symmetries, such as grand unified symmetries or supersymmetry,
or for radical extensions of confirmed physics, such as string
theory or extra dimensions~\cite{dob}.

\section{The ``generon'' model}
The model has two constituents and a confining group. The constituents are (1) the
 ``hexon'' ($h$) (from
Greek for six) a spin-1/2 fermion that
carries the color, lepton number, flavor and
helicity of one generation of quarks and leptons, and (2) the ``generon''
($g$) (from its role in constructing
the generations of quarks and leptons), a spin-0 boson that carries none of the standard
model quantum numbers. The confining group, 
the ``bindon'' group, is $[SU(3)\times SU(3)]_b$. These constituents 
and the confining group are the analogs 
for the present model of the quarks~\cite{gel,zwe} and the color degree of freedom~\cite{owgcolor} and the 
color $SU(3)$ group of the quark model.~\cite{nambucolor}
The basic particles of our composite model of quarks and leptons carry representations of
the bindon group. The hexon is a $(6,6)$ of
$[SU(3)\times SU(3)]_b$, and the generon is a $(3,3)$ of $[SU(3)\times SU(3)]_b$.

There are several motivations for the hexon to carry the degrees of
freedom of one generation of quarks and leptons: (1) this avoids strangeness changing 
neutral currents (FCNC), (2) this automatically satisfies the anomaly-matching conditions
proposed by G. 't Hooft~\cite{hoo} that are extremely difficult or impossible to satisfy
otherwise, (3) this allows a simple way to calculate the Cabibbo, Kobayashi, Maskawa (CKM)
matrix, (4) asymptotic freedom still holds, (5) the model is simple in contrast to
many composite models that are very complex. These good features come at the price of
a restricted goal, simply to understand why there are three generations of quarks and leptons that
differ only in their masses. We do not try to understand properties inside a single generation.

The hexon carries the quantum numbers of the standard model group, $SU(3)_c \times
SU(2)_w \times U(1)_{em}$. The fields in the hexon multiplet correspond to the fields in
a single generation of quarks and leptons of the standard model, including a right-handed
neutrino field. We could call them $h_u, h_d$, etc. corresponding to the $u$, $d$, etc.
members of a generation (in this example, the first generation). To save writing
subscripts we use the notation $h_u \equiv \mathbf{u}$, $h_d \equiv \mathbf{d}$, etc.
With this notation, the hexon has the 16 fields $\mathbf{u}_L^{\alpha},
\mathbf{d}_L^{\alpha}, \mathbf{\nu}_{eL}, \mathbf{e}_L, \mathbf{u}_R^{\alpha},
\mathbf{d}_R^{\alpha}, \mathbf{\nu}_{eR}, \mathbf{e}_R$, $\alpha = 1,2,3$. 
The Lagrangian for the standard model
group acting on these hexon fields (each of which is a $(6,6)$ of the new 
$[SU(3)\times SU(3)]_b$ confining bindon group) is the standard one, except that
the Yukawa terms that involve the Higgs are
\beee
\mathcal{L}_{Yukawa} = -G_{\mathbf{u}}[(\bar{L}_{\mathbf{q}} \bar{\phi})\mathbf{u}_R + hc] 
-G_{\mathbf{d}}[L_{\mathbf{q}} \phi)\mathbf{d}_R + hc]
 -G_{\mathbf{\nu}}[(\bar{L}_{\mathbf{l}} \bar{\phi})\mathbf{\nu}_R + hc] 
-G_{\mathbf{e}}[L_{\mathbf{q}} \phi)\mathbf{e}_R + hc].
\eeee
With
\beee
\begin{array}{cc}
\langle \phi \rangle_0 =
\left ( \begin{array}{c}
0\\
v/\sqrt{2}\\
\end{array}  \right )\\
\end{array}
\eeee
we find $m_{\mathbf{u}}=v G_{\mathbf{u}}/\sqrt{2}$, $m_{\mathbf{d}}=v G_{\mathbf{d}}/\sqrt{2}$,
$m_{\mathbf{\nu}}=v G_{\mathbf{\nu}}/\sqrt{2}$, $m_{\mathbf{e}}=v G_{\mathbf{e}}/\sqrt{2}$.
Note that the Cabibbo angle and the other parameters of the CKM matrix do not appear in these
Yukawa couplings. Quark and lepton mixings are accounted for in a new way described below.
%\beee
%\mathcal{L} = \mathcal{L}_{weakgauge} + \mathcal{L}_{stronggauge} + \mathcal{L}_{hl}
%+ \mathcal{L}_{hq} + \mathcal{L}_{Higgs},
%\eeee
%\beee
%\mathcal{L}_{stronggauge} = - \frac{1}{2} tr(G_{\mu \nu} G^{\mu \nu})
%\eeee
%\beee
%G_{\mu \nu}=\frac{1}{2} G_{\mu \nu}^l
%\lambda^l=\partial^{\nu}B_{\mu}-\partial^{\mu}B_{\nu}+ ig[B_{\nu},B_{\mu}],
%\eeee
%\beee
%B_{\mu}=\frac{1}{2}\lambda^l b_{\mu}^l
%\eeee
%\beee
%\mathcal{L}_{weakgauge} = - \frac{1}{4} F^l_{\mu \nu} F^{l~\mu \nu}
% - \frac{1}{4} f_{\mu \nu} f^{\mu \nu}
%\eeee
%\beee
%\mathcal{L}_{hl} = \bar{R}i\gamma^{\mu}(\partial_{\mu} 
%+ \frac{i g^{\prime}}{2} \mathcal{A}_{\mu} Y) R + 
%\bar{L}i\gamma^{\mu}(\partial_{\mu} 
%+ \frac{i g^{\prime}}{2} \mathcal{A}_{\mu} Y 
%+ \frac{i g}{2} \mathbf{\tau} \cdot \mathbf{b}_{\mu}) L,
%\eeee
%\beee
%\mathcal{L}_{hq} = \bar{\psi}(i \gamma^{\mu} \mathcal{D}_{\mu}-m)\psi
%\eeee
%\beee
%\begin{array}{cc}
%L_{hu,hd}
%=
%\left ( \begin{array}{c}
%\mathbf{u}_L\\
%\mathbf{d}_L\\
%\end{array}  \right )\\
%\end{array}
%\eeee
%\beee
%\begin{array}{cc}
%R_{hu,hd}
%=
%\left ( \begin{array}{c}
%\mathbf{u}_R\\
%\mathbf{d}_R\\
%\end{array}  \right )\\
%\end{array}
%\eeee
%\beee
%\begin{array}{cc}
%L_{h\nu,he}
%=
%\left ( \begin{array}{c}
%\mathbf{\nu}_L\\
%\mathbf{e}_L\\
%\end{array}  \right )\\
%\end{array}
%\eeee
%\beee
%\begin{array}{cc}
%R_{h\nu,he}
%=
%\left ( \begin{array}{c}
%\mathbf{\nu}_R\\
%\mathbf{e}_R\\
%\end{array}  \right )\\
%\end{array}
%\eeee
The hypercharge values are the usual ones.
%\beee
%\mathcal{L}_{Higgs}=
We choose the four Yukawa couplings very small so that they do not give significant masses
to the component fields of the hexon. We use the dynamical interaction of the hexon 
and the generons in the composite
state to give the main masses of the quarks and leptons.
Because of these choices, the model simultaneously accomplishes two
things. First, it removes the most contrived aspect of the standard model: the choice of Yukawa
couplings to tune the quark and lepton masses and mixings to their observed values. Secondly,
it decouples the relation between the quark and lepton masses and the decay rates of the
Higgs which follows from the Yukawa couplings. This means that in this model the Higgs mass,
its decay rates, and bounds on both the mass and the decays must be reconsidered 
systematically. This observation is due to Claus Montonen~\cite{mon}. We hope to join him in
doing this systematic re-evaluation of the properties of the Higgs meson in a subsequent paper.

Present experimental evidence, which is somewhat model dependent, 
requires that the compositeness scale be greater than 
$4$ to $9~TeV$~\cite{com}. We assume a compositeness scale of this order.

At high energies the standard model gauge group will probe inside the quarks and leptons and
will see the hexon and 
the constituent generation to which the hexon belongs. At low energies the standard model gauge
group will
see the three generations of quarks and leptons. There will be a complicated transition
region between these domains.
 
Coupling the hexon to confined scalars should preserve the approximate
chiral symmetry of the hexon that protects the masses of the composite states that correspond
to the quarks and leptons of the three generations from acquiring masses of the order of
the inverse of the characteristic distance scale of the bound states.

We require the generations to be $[SU(3)\times SU(3)]_b$ singlets that are
single-centered in the sense that they cannot be factored into the product of two
$[SU(3)\times SU(3)]_b$ singlets. There are exactly three composites that 
meet this requirement:
$I ~(h\bar{g}\bar{g})$, $II ~(h(gg)\bar{g})$ and $III ~(h(gg)(gg))$; we choose these as
the three generations of quarks and leptons~\cite{geo}.\footnote{We note that without the
constraint that the $b$'s are bosons there would be two independent $(hgggg)$ states.} 
We use the product group
$[SU(3)\times SU(3)]_b$ to bind the generons, rather than just one $SU(3)_b$ group, because
with only one such group the bound states with more than one boson in the 
same orbital state would
vanish; in particular the antisymmetric state of $(gg)$ in the $\bar{3}$ would vanish. 

The Lagrangian for the generon model has terms for the various constituents. There is a
standard set
of terms for the $[SU(3) \times SU(3)]_b$ gauge interactions of the hexons,
the generons and the bindons. This product group will lead to asymptotic freedom for
$n_f \leq 33$, where $n_f$ is number of fermions that couple to the bindons. There is the
term that produces generation mixing and changing,
\beee
\mathcal{L}_{mix}=\lambda \epsilon^{\alpha_1 \beta_1 \gamma_1} \epsilon^{\alpha_2 \beta_2 \gamma_2}
g_{\alpha_1,\alpha_2}g_{\beta_1,\beta_2}g_{\gamma_1,\gamma_2} +  hc.     \label{mix}
\eeee
This mixing term will produce first order transitions between neighboring generations, but not
between generations $I$ and $III$, so with choice of $\lambda$ we can understand why 
$I$-$III$ mixing is smaller than the mixing between neighboring generations (at least for
quarks). Calculations to be done in a subsequent paper will determine whether or not this
mixing term can be chosen to lead to the observed CKM matrix for quarks 
and the analogous mixing for neutrinos. 
%With $\lambda$
%complex the CKM matrix will contain a phase that parametrizes CP violation.
Because $\mathcal{L}_{mix}$ is a superrenormalizable interaction, radiative corrections to the
action of $\mathcal{L}_{mix}$ will be finite.
There are mass terms for the hexon and
generon. Finally there are the usual 
$SU(3)_{color} SU(2)_{weak} \times U(1)_{em}$ gauge fields
of the standard model which act on the hexon. We chose the Yukawa coupling constants that
determine the hexon masses via the Higgs mechanism to be so small that the hexon masses do
not contribute to the mass fits for the quarks and leptons that we give later.

\section{Generation mixing}
The states $I, II$ and $III$ are not the fermion mass eigenstates because $\mathcal{L}_{mix}$
causes transitions between the
generations. The mass eigenstates, $\Psi_u,\Psi_c,\Psi_t$
(and $\Psi_d,\Psi_s,\Psi_b$) will be related to the states
$I_u$, $II_u$, $III_u$ (and $I_d$, $II_d$, $III_d$)
by a $3 \times 3$ matrix for the up-type quarks,,
\beee
\begin{array}{ccc}
\left ( \begin{array}{c}
\Psi_u\\
\Psi_c\\
\Psi_t\\
\end{array}  \right )
=
M^{(u)}
\left ( \begin{array}{c}
I_u\\
II_c\\
III_t\\
\end{array}   \right )
\end{array}
\eeee
and an analogous matrix $M^{(d)}$ for the down-type quarks.
We assume that the weak currents couple to $h_u$ and $h_d$. Then the CKM matrix will be
$M^{(u)} M^{(d)\dagger}$. There will be analogous relations for the neutrino mixings.
  %that enter the currents that
%couple to the weak and electromagnetic gauge bosons. The states that correspond to the
%mass eigenstates of the fermions have mixings of the above states due to the term
%$\mathcal{L}_{mix}$ that produces transitions between generations and
%is responsible for the CKM quark mixings as well as the neutrino
%mixings.
Note that for $\mathcal{L}_{mix}$
the existence of two $SU(3)_b$ confining groups is crucial just as it is for the states that
represent the generations. 
%We have checked that for many calculations the doubling of the
%$SU(3)_b$ groups only changes a numerical factor, 
%so they don't both have to be kept, except when
%dropping one of them would result in an operator vanishing identically.

We write the detailed form of the three generations, suppressing
the indices carried by the hexon that account for the flavor,
color and chirality quantum numbers of a single generation, 
\beee
I: h_{\alpha_1 \beta_1,\alpha_2
\beta_2}\bar{g}^{\alpha_1,\alpha_2}\bar{g}^{\beta_1,\beta_2},
\eeee \beee II: h_{\alpha_1\beta_1,\alpha_2
\beta_2}\bar{g}^{\alpha_1,\alpha_2} \epsilon^{\beta_1 \gamma_1
\delta_1} \epsilon^{\beta_2 \gamma_2 \delta_2}
b_{\gamma_1,\gamma_2} b_{\delta_1,\delta_2}, \eeee \beee III:
h_{\alpha_1\beta_1,\alpha_2 \beta_2} \epsilon^{\alpha_1 \gamma_1
\delta_1} \epsilon^{\alpha_2 \gamma_2 \delta_2} \epsilon^{\beta_1
\epsilon_1 \zeta_1} \epsilon^{\beta_2 \epsilon_2 \zeta_2}
b_{\gamma_1, \gamma_2} b_{\delta_1, \delta_2} b_{\epsilon_1,
\epsilon_2} b_{\zeta_1, \zeta_2}, 
\eeee 
where $h$ is separately
symmetric in the indices on each side of the comma, so that it is
a $(6, 6)$ of the bindon group.

\section{Dark matter candidate}
Any electrically neutral particle is a candidate for dark matter. In this model the
$g\bar{g}$ bound states are particularly appealing as dark matter candidates, because as
bound states of scalars with no unconfined quantum numbers they will not have any of the
interactions of the standard model. They will, of course, interact with gravity. 
They will interact at the level
of the constituents of the quarks and leptons via exchange of bindons. We expect this
interaction to be weak and short-ranged.
The main decay will be to quarks and antiquarks
and via hadronization to hadrons. The $g\bar{g}$ bound states are different from
other dark matter candidates in having spin zero and being neutral under the flavor groups
of the standard model.
Many of the $h \bar{h}$ bound states carry electric charge. These are clearly not candidates for
dark matter. The neutral ones could contribute to dark matter; however because $h$
and $\bar{h}$ carry standard model quantum numbers they 
are likely to decay
more rapidly to quark-antiquark pairs than the $g\bar{g}$ states.

\section{Generic predictions for models of this type}
The most important characteristic of this type of model is that the standard model
group acts on the fermionic constituent of the composite system, in this case the 
hexon, rather than directly on the quarks and leptons. This fact has profound consequences
for the Higgs meson which will be discussed in a separate paper.

Any composite model in which the generations are the lowest orbital states that
are low in mass because of chiral symmetry will have orbital excitations that will
not be protected by chiral symmetry. These excitations will develop masses of the order of
the inverse of the size of the ground state, most likely in the several TeV region. Further,
these lowest excitations will have orbital angular momentum one, so that
the excitations of the three generations of spin-1/2 quarks and leptons of the standard model
will have spins 1/2 and 3/2 and opposite parity relative to the quarks and leptons.
Those particles that have strong interactions,
i.e. excited quarks, will be bound by the usual color interactions into
new hadrons. The decays of the excited quarks will induce decays of the new hadrons
via emission of pions and other mesons. The excited particles that have electromagnetic
interactions but no strong interactions, i.e. charged leptons,
will decay via emission of gamma rays; the particles, i.e. neutrinos, that have only
weak interactions will decay via emission of $W$'s and $Z$'s.

\section{Summary}
This model of three generations of quarks and leptons is simple. It has just two constituents. 
The first constituent, the hexon, $h$, is a spin-1/2 field that
carries the quantum numbers of one generation. The model accounts for generations by using
the second constituent, the generon, $g$, a scalar field, to construct 
the three generations as confined
bound states of the hexon and the generon and antigeneron.

The confining group is $[SU(3) \times SU(3)]_b$ with gauge bosons (bindons)
in the $(8, 8)$ representation of this
group. The only single-centered $[SU(3)\times SU(3)]_b$ singlets with one hexon are
$h\bar{g}\bar{g}$, $hgg \bar{g}$ and $hgggg$ which we identify with the three generations.
The model has a simple mechanism for mixing the
generations that will produce a CKM matrix. The model has 
predictions for higher mass particles, possibly in the TeV range.
Some of the higher-mass particles, particularly the heavy electron of $J^P 3/2^-$, will decay to 
an electron and a photon with no missing mass. This may be a unique such decay and will be searched
for at an early stage in the LHC program.~\cite{pc} Further the model has a dark matter candidate,
again possibly
in the TeV mass range.  We give exploratory comments about the mass
spectrum of the quarks and leptons in an Appendix in which the model accounts for the rapid 
increase in mass with generation by assuming an approximately
constant mass density due to the confined fields and configurations of the confined particles
that are quasi-one, quasi-two and quasi-three dimensional for the three generations. In this
way the three generations are identified with the three dimensions of space. The model can also
account for the different signs of the $u-d$ vs the $c-s$ and $t-b$ mass differences. 
\textit{All of
this discussion of masses is highly speculative and we do not believe that
the validity of this model should be based on our comments about the quark and lepton masses.}

\section{Future work}
In general we must develop this model from its present schematic form to a quantum field theory
just as the constituent quark model had to be developed into QCD. More specifically we must replace the
mixture of ideas based on naive non-relativistic physics, such as the fit to the quark and lepton
masses, to a coherent theory based on a specific Lagrangian. We have to justify that chiral symmetry
will keep the masses of the ground-state quarks and leptons light compared to the mass scale associated
with the inverse of the size of the bound states, without forcing the masses to be exactly zero.
Difficulties associated with doing this are discussed in~\cite{peccei}.
Another possibility
is to invoke a see-saw mechanism~\cite{seesaw} between a complete set of zero mass
ground state generations of quarks and leptons
and a set of $L^P=0^+, N=2$ excited generations at high mass. The see-saw mechanism would
then generate small masses for the known quarks and leptons. The very small masses of 
neutrinos could be
connected with the small matrix element between the ground-state neutrinos and the 
$L^P=0^+, N=2$ neutrinos. If the see-saw mechanism for neutrino masses
is realized in this way, this could give insight to why the neutrino mixings
differ so much from the quark mixings. One way or the other, we
must estimate the masses and mixings of
the quarks and leptons in the three generations. 
We must produce an argument, perhaps based on placing the three generations on branes of dimensions
1,2, and 3, for the factors of $L$ that we introduced in an ad hoc way. 
The superrenormalizable $\mathcal{L}_{mix}$
should allow a robust calculation of the CKM matrix. A general question to
explore is whether there is a qualitative impact of having two nested levels of confinement
as in this model.
We must study the impact of this model on the development of the
early universe when at very early times the hexons and generons might have been unbound.
The expected higher mass particles and their signatures at machines such as the LHC must be
worked out, as well as the ways in which the dark matter candidate, $g\bar{g}$, can be detected. 
Finally, we must work out the new scenario for the Higgs meson.

\section{Acknowledgements}
We thank Lay Nam Chang, Lew Licht, Rabi Mohapatra, Claus Montonen, Werner Nahm, 
Salah Nasri, Denjoe O'Connor, Dan-Olof Riska and Matts Roos for helpful
discussions. This work was supported in part by the National Science Foundation,
Grant No. PHY-0140301, by the Helsinki Institute of Physics, by the
Dublin Institute for Advanced Studies and by the Irish Fulbright Commision. 
It is a pleasure to thank
Dan-Olof Riska for his hospitality in Helsinki and to thank Denjoe O'Connor for
his hospitality in Dublin.

%\Appendix{Speculative ideas about the masses of the quarks and leptons}
\renewcommand{\theequation}{A-\arabic{equation}}
  % redefine the command that creates the equation no.
  \setcounter{equation}{0}  % reset counter 
  \section*{APPENDIX}  % use *-form to suppress numbering

The main elements that determine the masses of the particles in each generation are\\
(1) The volume in which the bindons are confined,\\
(2) The interaction between the hexon and the generons, and\\
(3) Group theory factors.\\
The volume in which the bindons are confined is determined by the confining interaction
between the hexon and the generons and antigenerons and the effective repulsion among 
the generons and antigenerons due to their zero-point motion 
which makes them separate from each other
as much as possible. For generation $I$ with two antigenerons their repulsion leads to a linear
quasi-one-dimensional configuration; for generation $II$ with three generons and antigenerons
their repulsion leads to a triangular quasi-two-dimensional configuration;
for generation $III$ with four generons their repulsion leads to a tetrahedal
quasi-three-dimensional configuration. We associate the dimensionless factor $L/a$, where
$L$ is the large linear scale of the generations and $a$ is small linear scale, with
the mass ratios between the generations, so that the mass scales of the successive generations
are proportional to $La^2, L^2 a$, and $L^3$, respectively. In short, we associate the three
generations of quarks and leptons with
the three dimensions of space. Our picture here is analogous to the bag model of hadrons
in which volume factors play a role in determining hadron masses~\cite{bag}.
Group theory (Clebsch-Gordan coefficients) will also play a role. Since
the constituent picture of the quarks and leptons is only a schematic one
these ideas can, at best,
give only a qualitative understanding of the masses of the particles in the generations.
A more accurate
picture must take into account a many-body description of the quarks and leptons in which
hexon-antihexon pairs, generon-antigeneron pairs and bindons enter.

We suggest the following binding scheme that can lead to the qualitative pattern of quark
and lepton masses. We take a potential model in the spirit of the constituent quark model
of hadrons. The Hamiltonian is the sum of kinetic energy terms for each constituent and
potential energy terms for the interactions between the constituents. We assume a confining
potential between each hexon, generon and antigeneron pair~\cite{grelip}. Based on the one-gluon
exchange interaction we take this confining interaction to be
\beee
H_I = -\sum_{i \leq j}\sum_{\alpha}F_i^{\alpha} F_j^{\alpha} V(|\mathbf{x}_i-\mathbf{x}_j|)
\eeee
even though we realize that the actual interaction will be more complicated. We give the
values of the Casimir $C^{(3)}_2$ and the interaction $-F_1 \cdot F_2 $ in the table below.
Note that because our group is $SU(3) \times SU(3)$ the Casimirs are the squares of the
Casimirs for just $SU)(3)$; i.e., $C^{(3)}_2(r,r)=[C^{(3)}_2(r)]^2$. We also use
$-F_1 \cdot F_2=1/2(F_1^2+F_2^2-C^{(3)}_2(state)$.
\begin{center}
\begin{tabular}{|l|l|r|}              \hline
\emph{State} & $C^{(3)}_2$ & $-F_1 \cdot F_2 $   \\ \hline
$(h\bar{g})_{(3,3)}$         & 16/9       & 50/9                \\
$(h\bar{g})_{(15,15}$        & 256/9      & -70/9                \\
$(hg)_{(8,8)}$               & 9          & 35/18            \\
$(hg)_{(10,10)}$             & 36         & -104/9            \\
$(gg)_{(3^{\ast},3^{\ast})}$ & 16/9       & 8/9              \\
$(gg)_{(6,6)}$               & 100/9      & -34            \\
$(g\bar{g})_{(1,1)}$         & 0          & 16/9              \\
$(g\bar{g})_{(8,8)}$         & 9          & -49/2             \\  \hline
%\caption{Two-body state, Casimir $C^{(3)}_2$, and interaction}
\end{tabular}\\
\end{center}
According to this table the confining interaction
is attractive for $(h\bar{g})_{(3,)}$, which we want, but repulsive for the
$(h\bar{g})_{(15,15}$ and $(hg)_{(10,10)}$ states, which we don't want.
At first sight this seems to rule out a confined state for generations
$II$ and $III$; however what is crucial is whether or not the interaction is attractive
or repulsive when one constituent is removed from the assumed bound state. Since the bound
states that correspond to the three generations are all $[SU(3) \times SU(3)]_b$ singlets,
removing a generon or an antigeneron will lead to a $(3^{\ast} \times 3^{\ast})-(3 \times 3)$
interaction which is always attractive. 

We assume that the zero-point fluctuations of the generons makes them separate from 
the hexon and each other. These repulsive fluctuations play the role of making the three
generations roughly one, two, and three dimensional for generations
$I$, $II$, and $III$, respectively, as described above. Molecular models that illustrate these
structures are $CO_2$, $NH_3$ (ignoring the difference between $g$ and $\bar{g}$) and
$CH_4$. These structures provide the framework inside
which the bindons, generon-antigeneron pairs and possibly hexon-antihexon pairs
live with some roughly constant energy
density, so that the masses of the three generations increase as $L a^2$, $L^2 a$ and $L^3$ as
stated above, just as the energy density of the gluons and quark-antiquark pairs give the
bulk of the masses of the hadrons in the quark model~\cite{bag}. Note that this volume 
effect does not occur in the quark model; for example the masses of mesons, 
which are two-body composites, do not have a ratio of volume factors relative to baryons,
which are three-body composites. The difference here is that the product of confining
interactions means that the gauge flux between a pair of generons carries other gauge
degrees of freedom and these additional gauge degrees of freedom interact between the 
different gauge fluxes.

The pattern of masses inside each
generation will depend on the interactions of the elements of the hexon with the generons
and antigenerons. If we take the hexon to be a $16$ of an $SO(10)$ multiplet, then the
up-down splitting in each generation will depend on the different interactions of
the $h_u$ and $h_d$ of the hexon $16$ with the generons and antigenerons. Thus if
the $h_u-\bar{g}$ and $h_d-\bar{g}$ interactions have the opposite sign to the
$h_u-g$ and $h_d-g$ (or the $h_u-(gg)$ and $h_d-(gg)$) interactions,
then we can understand why the $u-d$ mass difference
reverses between generation $I$ and generations $II$ and $III$.

In order to see if inserting different powers of $L$ for the three generations seems
reasonable, we made a exploratory fit to the six quark masses using parameters for
the $(h_ugg)$, $(h_dgg)$, $(h_u\bar{g})$, $(h_d\bar{g})$ interactions and the $g$ mass together
with a sixth parameter $a/L$. We introduce the parameters $(h_ugg)$, and $(h_dgg)$ because we
view the $gg$ pairs that couple to the $h$ as being in a $\bar{3} \times \bar{3}$ of
$[SU(3) \times SU(3)]_b$. We used the formulas
\beee
m_u = 2  (h_u\bar{g}) La^2 + 2 m_g,
\eeee
\beee
m_d = 2  (h_d\bar{g}) La^2 + 2 m_g,
\eeee
\beee
m_c = (h_u\bar{g}) L^2a + (h_ugg) L^2a + 3 m_g,
\eeee
\beee
m_c = (h_d\bar{g}) L^2a + (h_dgg) L^2a + 3 m_g,
\eeee
\beee
m_t = 2  (h_ugg) L^3a + 4 m_g,
\eeee
\beee
m_b = 2  (h_dgg) L^3a + 4 m_g.
\eeee
We used the constituent quark masses given by Scadron, 
Delburgo and Rupp~\cite{scad} as input masses. 
 We found that the quark masses scaled by the appropriate
factors of $a/L$ are all of comparable order of magnitude, 
and differ much less than the quark masses,
 and regard this as an indication
that the idea of inserting different powers of $L$ in the formulas for the quark
masses is worth pursuing. We found the ratio $a/L \sim 1.6 \times 10^{-2}$ or, equivalently,
$L/a \sim 63$. We assume
$a \ll L < \hbar/Mc$. We take $M$ to have a lower bound in the multi-TeV region~\cite{com}.
The result, with numerical parameters in $MeV/c^2$, is
\begin{center}
\begin{tabular}{|l|r|}              \hline
\emph{Parameter} & Value   \\ \hline
$h_{\mathbf{u}\bar{g}} a^3$ & $-8.53 \times 10^{-2}$       \\
$h_{\mathbf{d}\bar{g}} a^3$ & $-6.53 \times 10^{-3}$       \\
$h_{\mathbf{u}gg} a^3$   & $3.45  \times 10^{-1}$       \\
$h_{\mathbf{d}gg} a^3$   & $7.92  \times 10^{-3}$       \\
$a/L$           & $1.58  \times 10^{-2}$       \\
$m_g$           & $1.72  \times 10^2$          \\  \hline
\end{tabular}
\end{center}

We made the analogous calculation to fit the lepton masses. We used the charged lepton
masses given in~\cite{pdg} and took the neutrino masses to be zero for this exploratory
calculation. For the leptons,
there are two possible values for $a/L$. We give the results for both,
with numerical parameters again in $MeV/c^2$.
\begin{center}
\begin{tabular}{|l|r|}              \hline
\emph{Parameter} & Value   \\ \hline
$h_{\mathbf{\nu}\bar{g}} a^3$ & $0$       \\
$h_{\mathbf{e}\bar{g}} a^3$ & $6.34 \times 10^{-4}$       \\
$h_{\mathbf{\nu} gg} a^3$   & $0$       \\
$h_{\mathbf{e}gg} a^3$   & $1.35  \times 10^{-5}$       \\
$a/L$           & $2.48  \times 10^{-3}$       \\
$m_g$           & $0$          \\  \hline
\end{tabular}
\end{center}
The second fit is
\begin{center}
\begin{tabular}{|l|r|}              \hline
\emph{Parameter} & Value   \\ \hline
$h_{\mathbf{\nu}\bar{g}} a^3$ & $0$       \\
$h_{\mathbf{e}\bar{g}} a^3$ & $2.97 \times 10^{-2}$       \\
$h_{\mathbf{\nu} gg} a^3$   & $0$       \\
$h_{\mathbf{e}gg} a^3$   & $1.40  $       \\
$a/L$           & $1.16  \times 10^{-1}$       \\
$m_g$           & $0$          \\  \hline
\end{tabular}
\end{center}
We plan to study other schemes to generate the quark and lepton masses in the composite states
in collaboration with A.L. Licht in a later paper.

\end{document}